\documentclass[fleqn,usenatbib]{mnras}

\usepackage{newtxtext,newtxmath}

\usepackage[T1]{fontenc}

\DeclareRobustCommand{\VAN}[3]{#2}
\let\VANthebibliography\thebibliography
\def\thebibliography{\DeclareRobustCommand{\VAN}[3]{##3}\VANthebibliography}
\newcommand{\fermi}{{\it Fermi-LAT} }
\newcommand{\gray}{$\gamma$-ray }
\newcommand{\grays}{$\gamma$-rays }
\newcommand{\es}{1ES 1218+304 }

\usepackage{graphicx}	
\usepackage{amsmath}	
\usepackage{amssymb}	

\title[Multiwavelength emission from 1ES 1218+304]{Broadband Study of High-Synchrotron-Peaked BL Lac Object 1ES 1218+304}

\author[N. Sahakyan]{
N. Sahakyan,$^{1,2}$\thanks{E-mail: narek@icra.it}
\\
$^{1}$ICRANet-Armenia, Marshall Baghramian Avenue 24a, Yerevan 0019, Armenia\\
$^{2}$ICRANet, P.zza della Repubblica 10, 65122 Pescara, Italy
}

\date{Accepted XXX. Received YYY; in original form ZZZ}

\pubyear{2020}

\begin{document}
\label{firstpage}
\pagerange{\pageref{firstpage}--\pageref{lastpage}}
\maketitle

\begin{abstract}
The origin of the multiwavelength emission from the high-synchrotron-peaked BL Lac 1ES 1218+304 is studied using the data from {\it Swift UVOT/XRT}, {\it NuSTAR} and {\it Fermi-LAT}. A detailed temporal and spectral analysis of the data observed during 2008-2020 in the $\gamma$-ray ($>100$ MeV), X-ray (0.3-70 keV), and optical/UV bands is performed. The $\gamma$-ray spectrum is hard with a photon index of $1.71\pm0.02$ above $100$ MeV. The {\it Swift UVOT/XRT} data show a flux increase in the UV/optical and X-ray bands; the highest $0.3-3$ keV X-ray flux was $(1.13\pm0.02)\times10^{-10}{\rm erg\:cm^{-2}\:s^{-1}}$. In the 0.3-10 keV range the averaged X-ray photon index is $>2.0$ which softens to $2.56 \pm 0.028$ in the 3-50 keV band. However, in some periods, the X-ray photon index became extremely hard ($<1.8$), indicating that the peak of the synchrotron component was above $1$ keV, and so 1ES 1218+304 behaved like an extreme synchrotron BL Lac. The hardest X-ray photon index of 1ES 1218+304 was $1.60 \pm 0.05 $ on MJD 58489. The time-averaged multiwavelength spectral energy distribution is modeled within a one-zone synchrotron self-Compton leptonic model using a broken power-law and power-law with an exponential cutoff electron energy distributions. The data are well explained when the electron energy distribution is $E_{\rm e}^{-2.1}$ extending up to $\gamma_{\rm br/cut}\simeq(1.7-4.3)\times10^{5}$, and the magnetic field is weak ($B\sim1.5\times10^{-2}$ G). By solving the kinetic equation for electron evolution in the emitting region, the obtained electron energy distributions are discussed considering particle injection, cooling, and escape.
\end{abstract}




\section{Introduction}
Active galactic nuclei (AGNs) with a bolometric luminosity of up to $10^{48}\:{\rm erg\:s^{-1}}$ are the most powerful non-explosive sources in the Universe. Among AGNs, blazars are the most extreme class dominated by nonthermal emission extending from radio to Very High Energy (VHE; $>$ 100 GeV) \gray band. The blazar features are best described when assuming that the relativistically moving plasma in the jet is closely aligned with the line of sigh of the observer \citep{1995PASP..107..803U}. The observations in various bands provide different windows on blazar physics, allowing to investigate the accretion disc, innermost jet (sub-parsec) as well as the knots and hotspots of large-scale jets. Most recently, the observation of VHE neutrinos from TXS 0506+056 \citep{2018Sci...361..147I, 2018Sci...361.1378I} opened another window for studying the physics of blazar jets. Combination of electromagnetic and neutrino observations could provide most detailed information on the physics at work in the jets (e.g., for TXS 0506+056 \citep{2018ApJ...863L..10A,2019NatAs...3...88G, 2019MNRAS.483L..12C, 2018ApJ...864...84K, 2018ApJ...865..124M, 2018arXiv180705210L,2018arXiv180900601W, 2018MNRAS.480..192P, 2018ApJ...866..109S, 2019MNRAS.484.2067R} and \citep{2019A&A...622A.144S}).\\
Commonly, blazars are divided into two subclasses: flat-spectrum radio quasars (FSRQs) and BL Lacertae objects \citep{1995PASP..107..803U}. The optical spectrum of FSRQs reveals strong broad emission lines, while that of BL Lacs has weak or no lines. The spectral energy distribution (SED) of blazars in $\nu F\nu$ representation has two components \citep[e.g.,][]{2017A&ARv..25....2P} and is characterized by two broad peaks: the low energy component commonly explained by synchrotron emission of relativistic electrons, peaks between the IR and the X-ray bands. When the synchrotron peak ($\nu_{s}$) is $\nu_{s}<10^{14}$ Hz in the rest-frame, blazars are called low synchrotron peaked (LSP) sources, and when $10^{14}<\nu_{s}<10^{15}$ Hz and $\nu_{s}>10^{15}$ Hz are intermediate synchrotron peaked (ISP) and high synchrotron peaked (HSP) sources, respectively \citep{1994MNRAS.268L..51G, 2010ApJ...716...30A}. In this division, FSRQs are almost exclusively LSPs.\\
There are various models explaining the second peak in the SED. In the leptonic scenarios, this is explained as inverse Compton (IC) scattering of photons provided by the synchrotron emission of the jet itself \citep[i.e., synchrotron self Compton (SSC)][] {1992ApJ...397L...5M, bloom, ghisellini} or produced external to the jet \citep{sikora,1994ApJS...90..945D}. The most widely used sources of external seed photons are disc photons reflected from broad line region (BLR) clouds \citep{sikora} or IR photons emitted from the dusty torus \citep{2000ApJ...545..107B, 2009MNRAS.397..985G}. Since the BLR lines are weak or absent in BL Lacs their SEDs are usually modeled using SSC while those of FSRQs by external IC mechanism. In the alternative hadronic scenarios, the second component is modeled by proton synchrotron emission 
\citep[e.g.,][]{2001APh....15..121M}, photopion production \citep{1993A&A...269...67M, 1989A&A...221..211M, 2001APh....15..121M, mucke2, 2013ApJ...768...54B} or $pp$ interaction \citep{dar, araudo10, bednarek15, beall, bednarek97}.\\
The synchrotron peak location is defined by the maximum energy at which the electrons are accelerated. In this context, HSPs are not the highest-energy end of the blazar sequence, and \citet{2001A&A...371..512C} found objects with a hard synchrotron X-ray spectrum of at least up to $\sim100$ keV. These extreme synchrotron BL Lacs or extreme HSPs (EHSPs) show a synchrotron peak energy above $2.4\times 10^{17}$ Hz (1 keV), an order of magnitude higher than that of standard HSPs. For example, during the flares of Mkn 501 the synchrotron peak  reached $\sim100$ keV \citep{1998ApJ...492L..17P}. Due to this shift, in the optical band the emission from EHSPs is generally dominated by the thermal emission of the giant elliptical host galaxy. The radio properties of EHSPs are in general similar to those of HSPs but rather with a low flux. In addition to extreme synchrotron BL Lacs, there are BL Lacs extreme in \grays which after extragalactic background light (EBL) correction demonstrate a very hard intrinsic photon index of up to and beyond 1 TeV \citep{2015MNRAS.451..611B,2011MNRAS.414.3566T}. There is no clear relation between extreme synchrotron and TeV blazars and these two extreme behaviors should not necessarily appear together. Hard spectral photon indexes above 1 TeV due to similar hard index of the emitting particles represent major difficulties for current particle acceleration and emission theories. These extreme blazars are also discussed as possible sources of VHE neutrinos and cosmic rays \citep{2016MNRAS.457.3582P, 2017MNRAS.468..597R}.\\
The hard High Energy (HE; $>100$ MeV) \gray spectrum of HSPs implies that particles are efficiently accelerated up to VHEs in their jets, so their detailed study is interesting from the theoretical point of view. One of such HSPs, is \es at $z=0.182$ \citep{2003A&A...412..399V} which has been for the first time observed at VHEs by {\it MAGIC} \citep{2006ApJ...642L.119A} and then by {\it VERITAS} telescopes \citep{2009ApJ...695.1370A}. The observed $\sim160$ GeV and $\sim1.8$ TeV emission is described with a hard \gray photon index of $1.86 \pm 0.37$ after EBL correction \citep{2009ApJ...695.1370A}. Next, the {\it VERITAS} observations during the active state in 2009 provided the first evidence of variability of VHE \gray emission of \es with a flux doubling time scale of $\leq1$ day \citep{2010ApJ...709L.163A}. In the HE \gray band, \es appears with a hard photon index of $1.72\pm0.02$, as observed by Fermi Large Area Telescope (\fermi) \citep{2020ApJS..247...33A}, with the emission extending beyond 100 GeV well in agreement with the data in the VHE \gray band. \es was identified as an X-ray source in the early observations \citep{1979MNRAS.187..109W} and since then it was always observed with X-ray telescopes. Considering the unusually hard VHE \gray spectra of \es for its redshift, its observations were also used to constrain the EBL absorption density \citep[e.g.,][]{2020A&A...633A..74K} or extragalactic magnetic field \citep{2011A&A...529A.144T}.\\
The multiwavelength observations of \es over years provided a large amount of data in different bands. First, more than eleven years of \fermi observations will allow detailed temporal and spectral analyses of \gray data which combined with {\it MAGIC/VERITAS} data provides the \gray spectrum in the large energy range from 100 MeV to $\sim1$ TeV. Moreover, using the new PASS 8 event selection and instrument response function, the spectrum can be investigated with improved statistics at higher energies, which is crucial for identifying the peak of the HE component. Frequent observations of \es with Neil Gehrels Swift Observatory (\cite{2004ApJ...611.1005G}, hereafter {\it Swift}) provided unprecedented data both in the optical/UV and X-ray bands, allowing to perform a detailed investigation of the flux variation in these bands. This broadband coverage allows to constrain the SED of \es in different periods, which is then used for theoretical modeling. Together with {\it Swift}, the {\it NuSTAR} observation will shape the peak of the low energy component, which in turn allows to derive the main parameters characterizing the jet of \es (emitting electron distribution, magnetic field, jet power, etc.). \es belongs to the group of blazars that exhibit hard \gray spectrum from MeV/GeV to TeV band, which implies the emission is most likely produced from fresh accelerated electrons allowing to test various acceleration and cooling processes for the emitting particles. The combination of this with the available data, makes \es an ideal target for exploring the physics of blazar jets.\\
The purpose of this paper is to investigate the origin of broadband emission from \es by analyzing the most recent available data. In Section \ref{sec1} the \gray data extraction and analyses are presented and discussed while X-ray and optical data analyses are in Section \ref{sec2}. The origin of broadband emission as well as the SED modeling are given in Section \ref{sec3}. The time dependent formation of emitting electron spectrum is discussed in Section \ref{sec4}. The discussions and conclusions are given in Section \ref{sec5}.\\

\section{Fermi-LAT data extraction and analyses}\label{sec1}
\fermi is a pair-conversation telescope sensitive to $>100$ MeV \grays \citep{2009ApJ...697.1071A}. By default it operates in the survey mode scanning the entire sky every three hours. Operating since 2008, \fermi has provided a most detailed and deeper view of the \gray sky.\\
In the current study, \gray data from the observation of \es from August 2008 to April 2020 were obtained from the data portal and analyzed using the standard analysis procedure provided by the \fermi collaboration. The events in the energy range from 100 MeV to 600 GeV within a circular region of $11^{\circ}$ radius centered on the \gray position of \es were analyzed using  Fermi ScienceTools (1.2.1) package with P8R2\_SOURCE\_V6 instrument response functions. A zenith angle cut of $90^{\circ}$ was applied to reduce the contamination due to the \grays from the Earth's limb. The model file containing the spectral parameters of all known \gray emitting sources located within a $11^{\circ}$+$5^{\circ}$ region was generated by {\it make4FGLxml.py} script based on the fourth \fermi source catalog of \gray sources (4FGL) \citep{2020ApJS..247...33A}. The Galactic and extragalactic diffuse \gray emission was parametrized using {\it gll\_iem\_v07} and {\it iso\_P8R3\_SOURCE\_V2\_v1} models. The parameters of all sources within the $11^{\circ}$ region around \es as well as the normalization of diffuse components were left free to vary during the fitting while the spectral parameters of all other sources were fixed to their values given in the 4FGL.\\
The data from a $15.5^{\circ}\times15.5^{\circ}$ square region are divided into a spatial pixel size of $0.1{^\circ} \times 0.1^{\circ}$ and into 38 logarithmically equal energy bins. The best match between the model and the data is obtained by the binned likelihood analysis method implemented in {\it gtlike} tool. In the considered $\sim11.7$ years, \es is detected with an overall significance of $77.2\sigma$ ($\sigma=\sqrt{TS} $ where $TS = 2 (log L_1 - log L_0)$ and $L_{1}$ and $L_{0}$ are  the maximum likelihood values obtained when fitting the observed data using the null and alternative hypotheses, respectively). The best fit results a relatively hard \gray photon index of $1.71\pm0.02$ with a \gray flux of $(1.89\pm0.09)\times10^{-8}{\rm photon\;cm^{-2}\;s^{-1}}$ in the energy range from 100 MeV to 600 GeV. The SED of \es generated by running the {\it gtlike} tool separately for ten energy bands is shown in Fig. \ref{sed}.\\
\begin{figure*}
   \centering
   \includegraphics[width=\hsize]{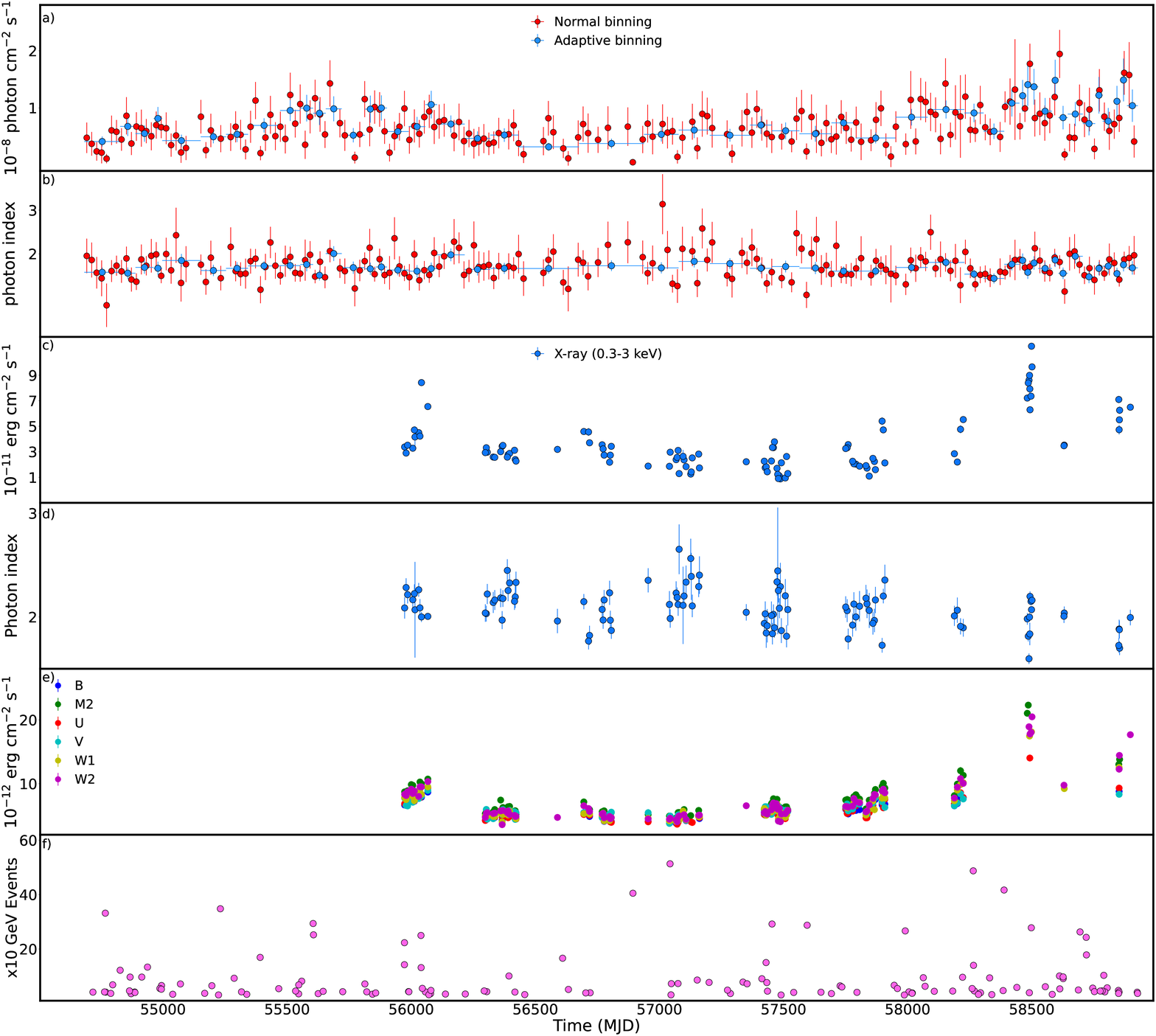}
      \caption{Multiwavelength light curve of \es during 2008-2020. {\it a)} and {\it b)} \gray flux and photon index computed for normal and adaptively time binning. {\it c} and {\it d} {\it Swift XRT} measured X-ray flux and photon index variation in time. {\it e)} {\it Swift UVOT} measured UV/optical fluxes in V, B, U, W1, M2, and W2 bands. {\it f)} The arrival time of HE photons from the direction of \es.}
         \label{multilight}
   \end{figure*} 
The \gray light curve is generated to investigate the flux and photon index variation in time. The $>500$ MeV events were only considered in the unbinned likelihood analyses, since, due to the hard spectrum of \es during short periods the number of photons is not enough at lower energies. The model file obtained from the binned likelihood analyses was used for the light curve calculations fixing the photon indexes of all background sources allowing only their normalization to vary. The normalization of both background models was fixed as no variability is expected from them. When the source detection significance is $TS<4$, only upper limit is computed.\\
Fig. \ref{multilight} panels {\it a)} and {\it b)} show the change of the \gray flux and photon index calculated for $20$ day intervals, respectively. Despite the increase of the \gray flux in some periods, no high-amplitude flares are observed. This is in agreement with the results of 4FGL where \es was flagged as variable source \citep[i.e., the variability index over two-month intervals is $95.6$ ][]{2020ApJS..247...33A}. The hard \gray photon index of \es implies that the emission is mostly at higher energies where the number of observed photons is low, so no comprehensive variability studies (e.g., short time scale variation) are possible. Next, using the adaptive binning algorithm \citep{2012A&A...544A...6L}, the \gray light curve is computed. In this method, the time bins have been optimized to have a fractional uncertainty of 20 \% above the optimal energy of $E_{\rm opt}=394.1$ MeV \citep[for the calculation of $E_{\rm opt}$ see][]{2012A&A...544A...6L}. The light curve generated by this strategy allows us to search variability, which is sometimes not visible in the regular time binning \citep{2013A&A...552A..11R, 2017MNRAS.470.2861S, 2017ApJ...848..111B, 2017A&A...608A..37Z, 2018A&A...614A...6S, 2018ApJ...863..114G, 2013A&A...557A..71R}. The flux and photon index calculated by adaptive binning method shown in light blue correspondingly in Fig. \ref{multilight} a) and b). The photon index is relatively stable, being always below $2.0$, which is natural, as HSPs usually have a hard \gray photon index but the \gray flux is sometimes above its average level. Despite large uncertainties, an increases in the flux had been observed around MJD 55578, 55879, 56080 and after 58000. For example, on MJD $58869.84\pm13.82$ the flux above $E_{\rm opt}=394.1$ MeV  was $(1.49\pm0.37)\times10^{-8}{\rm photon\;cm^{-2}\;s^{-1}}$ with a photon index of $1.75\pm0.15$, with a $10.2\sigma$ detection significance. It corresponds to a flux of $(3.89\pm0.84)\times10^{-8}{\rm photon\;cm^{-2}\;s^{-1}}$ above 100 MeV. Or on MJD $58594.42\pm12.95$ and $56080.57\pm23.54$, the \gray flux and photon index were $(5.15\pm1.05)\times10^{-8}{\rm photon\;cm^{-2}\;s^{-1}}$ and $1.86\pm0.16$, and $(2.87\pm0.55)\times10^{-8}{\rm photon\;cm^{-2}\;s^{-1}}$ and $1.66\pm0.13$, respectively. Therefore, even if the flux is above its average level ($\sim2$ times), the photon index does not change substantially. However, in normal binning, the hardest photon index of $1.30\pm0.17$ was observed from MJD 58262 to 58282 when the detection significance was $10.1\sigma$. The spectrum in this period was investigated further. The source emission above $1$ GeV can be described by the $1.39\pm0.16$ photon index and the emission extending up to $\sim200$ GeV with a flux of $(5.09\pm1.47)\times10^{-9}{\rm photon\;cm^{-2}\;s^{-1}}$.\\
\es is also a source of VHE photons due to its relatively hard photon index. Using the output model file obtained after running {\it gtlike}, with the {\it gtsrcprob} tool, the probability of VHE events from the direction of \es is computed. The distribution of highest-energy events ($>30$ GeV) is presented in Fig. \ref{multilight} (f). Interestingly, there are many $> 100$ GeV photons within the inner region around \es with a high probability of being associated with it. For example, the $169.2$, $178.4$ and $487.4$ GeV events with probabilities of $0.99996$, $0.99993$ and $0.99988$, respectively, were observed within a circle of $0.015^{\circ}$, $0.024^{\circ}$ and $0.035^{\circ}$, respectively. The highest energy event of $513.2$ GeV has been detected on MJD 57042.8 within a circle of $0.18^{\circ}$ around \es with the $0.99496$ probability to be associated with it.
\section{X-ray and optical/UV observations}\label{sec2}
The X-ray emission from \es is investigated by analyzing the data collected by {\it Swift XRT} \citep{2004ApJ...611.1005G} and {\it NuSTAR} \citep{2013ApJ...770..103H}. This provides X-ray data in a large energy range of $0.3-70$ keV which is important, as for HSPs this range corresponds to the highest energy tail of the low energy component.
\begin{figure}
   \centering
   \includegraphics[width=0.48 \textwidth]{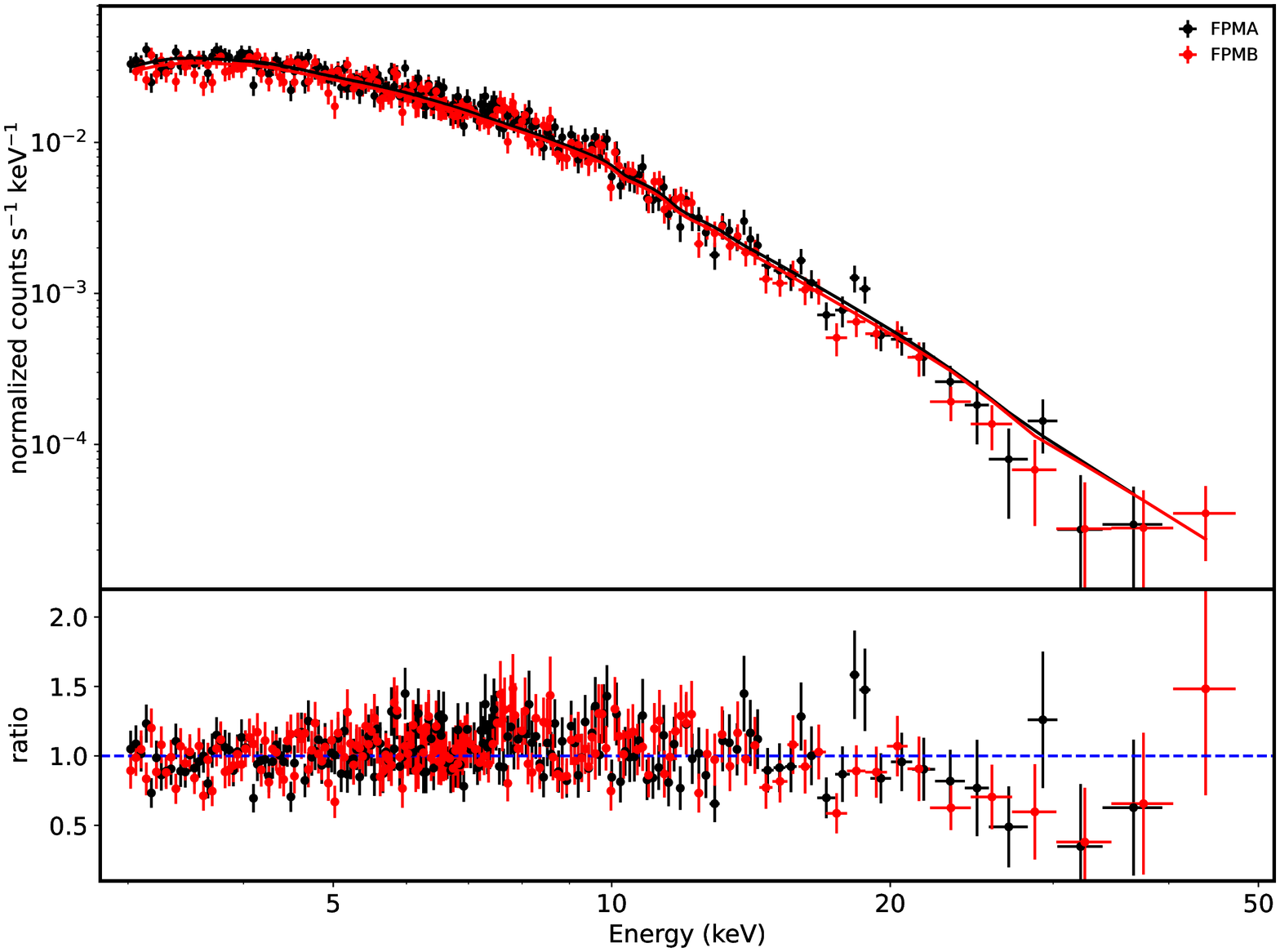}
      \caption{{\it Upper panel:} {\it NuSTAR} spectra (FPMA in black and FPMB in red) and the best fit model. {\it Lower panel:} the ratios (data/model) for power-law model.}
         \label{nustar}
   \end{figure} 
\subsection{{\it Swift XRT}}\label{ss}
\begin{figure*}
   \centering
   \includegraphics[scale=0.65]{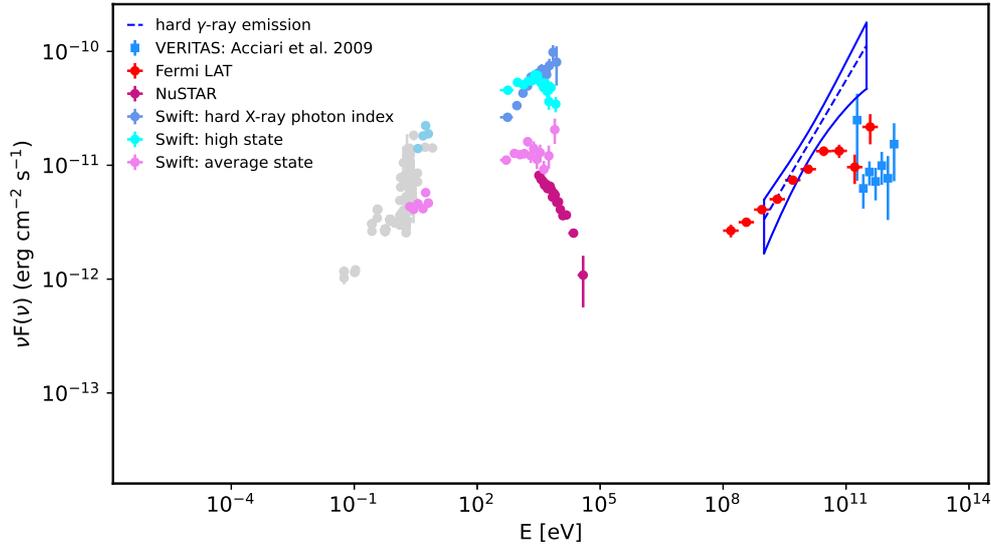}
      \caption{Multiwavelength SED of \es for different periods. The red data corresponds to the \fermi spectrum averaged over 11.7 years, and the blue bowtie shows the spectrum during the hard emission period. The {\it UVOT} data in light blue corresponds to the highest flux in U, W1, M2 and W2 filters observed around MJD $58490$. The archival data from SSDC are in gray. VHE \gray data from the {\it VERITAS} observation are in light blue squares.}
         \label{sed}
   \end{figure*}

{\it Swift} observed \es 116 times between 2008 and 2020. All {\it XRT} data are analyzed using the {\it Swift} XRTPROC pipeline, which is an automatic script for downloading and analyzing {\it XRT} data. The script first presented in \citet{2015JHEAp...7..173G} and further updated in the context of the Open Universe initiative \citep{2018arXiv180508505G}, is based on the official XRT Data Analysis Software (XRTDAS). For the source region, photons were counted over a circular region of 47 arcsec ($20$-pixel) radius centered on the source position, while for the background region a larger annulus was used, with inner and outer radii of 120 and 200 arcsec, correspondingly, centered on the source and selected to avoid any contaminating sources. The count rate in some observations was above $0.5$ ${\rm count\:s^{-1}}$, so the data is significantly affected by the pileup in the inner part of the point-spread function \citep{2005SPIE.5898..360M}. This effect was removed by excluding the events within a $3$-pixel radius circle centered on the source position. In this case, the source count selection radius was increased to 70 arcsec. The individual spectra were fitted with XSPEC12.10.1 adopting an absorbed power-law and log-parabola models, applying Cash statistics on ungrouped data.\\
The 0.3-3, 0.5-2, 2-10, 3-7 keV X-ray fluxes as well as the 0.3-10 keV photon index are computed for each observation. In the X-ray band, the flux gradually increases around MJD 58500 with the highest 0.3-3 keV X-ray flux of $(1.13\pm0.02)\times10^{-10}\:{\rm erg\:cm^{-2}\:s^{-1}}$ on MJD 58499.1, which is by a factor of $\sim5.6$ higher than the mean X-ray flux ($(2-3)\times10^{-11}\:{\rm erg\:cm^{-2}\:s^{-1}}$). It should be noted that from MJD 58482 to 58501, \es was in an active X-ray emission state, when the 0.3-3 keV flux changed in the range of $(6.29-9.65)\times10^{-11}\:{\rm erg\:cm^{-2}\:s^{-1}}$. Similar increases are also observed in the other considered intervals for the flux computation. The X-ray photon index varies as well, being of the order of $(2.1-2.2)$ for most of the time, but on MJD 58489.1 it was $1.60\pm0.05$ which is the hardest index recorded for this source (corrected for pile-up effect). There are thirty-four periods with a photon index of $<2.0$ and six with $<1.8$. Fig. \ref{sed} shows the {\it XRT} spectra when \es was in a bright (Obsid 30376106), moderately bright (Obsid 30376101), and average (Obsid 30376042) X-ray emitting states, as well as when the X-ray emission is with the hardest X-ray photon index (Obsid 30376099).  
\subsection{{\it NuSTAR}}
{\it NuSTAR} is a hard X-ray focusing satellite comprised of two Focal Plane Modules (FPMs): {\it FPMA} and {\it FPMB}, providing continuous coverage over a broad bandpass of 3-78 keV \citep{2013ApJ...770..103H}. \es was observed by {\it NuSTAR} on MJD 57349 with a net exposure of $49.5$ ksec. The {\it NuSTAR} data were processed with the NuSTARDAS software package available within HEASOFT package using the latest version of the calibration database (CALDB). The event file is cleaned and calibrated using {\it nupipeline} tool. The spectra of \es in the energy range of 3-79 keV is extracted from a circular region of 50$''$ radius centered at the source position whereas the background counts are extracted from a circle of 80$''$ from a nearby region on the same chip and avoiding source contamination. The spectra were binned so as to have at least 30 counts ${\rm bin}^{-1}$ and fitted assuming an absorbed power-law model.\\
Initially, the energy range from 3 to 79 keV have been considered for the fit. However, the count rate rapidly decreases above 50 keV and the background starts to dominate. Thus, the source parameters are estimated in the energy range 3-50 keV. The best fit resulted in $\Gamma_{\rm X-ray} = 2.56\pm 0.028$ and $F_{\rm X-ray}=(1.21\pm0.02)\times10^{-11}\: {\rm erg\:cm^{-2}\:s^{-1}}$ with $\chi^2/{\rm d.o.f}=1.04$ for 364 degree of freedom. The spectra of FPMA and FPMB are shown in Fig. \ref{nustar} with black and red, respectively. The high energy tail of X-ray spectra cannot be fitted satisfactorily by a simple power-law, and the model deviates from the data. Thus, an additional fit with a log-parabola is performed. The best-fit spectral parameters of the log-parabola fit are: $\alpha=2.22\pm0.10$, $\beta=0.45\pm0.13$ and $F_{\rm X-ray}=(1.14\pm0.02)\times10^{-11}\: {\rm erg\:cm^{-2}\:s^{-1}}$ with $\chi^2/{\rm d.o.f}=0.94$. 
\subsection{{\it Swift UVOT}}
\es was also observed with the {\it UVOT} instrument of {\it Swift} observatory. {\it UVOT} provides observations in three optical (V,B,U) and three UV (W1,M2,W2) filters \citep{2005SSRv..120...95R}. All the data available from the observations of \es were downloaded from the Swift archive and reduced using the HEAsoft version 6.26 with the latest release of of HEASARC CALDB. Source counts were extracted using a 5$''$  radius for all single exposures and all filters, while the background was estimated from different positions from a region with 20$''$ radius not being contaminated with any signal from the nearby sources. {\it uvotsource} tool was used to derive the magnitudes which were converted to fluxes using the conversion factors provided by \cite{poole}. Then, the fluxes were corrected for extinction using the reddening coefficient $E(B-V)$ from the Infrared Science Archive \footnote{ http://irsa.ipac.caltech.edu/applications/DUST/}.\\
The flux measured for all six filters (V, B, U, W1, M2, and W2) is sown in Fig. \ref{multilight} e). The light curve shows that like in the X-ray band, also the optical/UV flux shows few active periods. In the average state the flux in all filters is around $(3-5.5)\times10^{-12}\:{\rm erg\:cm^{-2}\:s^{-1}}$, which around MJD 56035, 57870 and 58200 moderately increases up to $\sim(8-9)\times10^{-12}\:{\rm erg\:cm^{-2}\:s^{-1}}$. Strong brightening of the optical/UV flux was observed after MJD 58482 when the flux reached  $\simeq10^{-11}\:{\rm erg\:cm^{-2}\:s^{-1}}$;  the absolute highest fluxes of $(2.23\pm0.05)\times10^{-11}\:{\rm erg\:cm^{-2}\:s^{-1}}$ and $(2.05\pm0.04)\times10^{-11}\:{\rm erg\:cm^{-2}\:s^{-1}}$ were observed in M2 and W2 filters on MJD 58486.10 and 58501.20, respectively. Fig. \ref{sed} shows the {\it UVOT} spectra for the average state (Obsid 30376044, pink) state and when the highest flux in U, W1, M2 and W2 filters were observed around MJD $58490$ (light blue).
\section{The Origin of Broadband Emission}\label{sec3}
The multiwavelength dataset analyzed in this paper allows to build broadband SEDs of \es in different emission states. The resulting broadband SED of \es is shown in Fig. \ref{sed}, displaying the standard double peaked structure. Archival data are extracted from SSDC SED Builder tool \footnote{https://tools.ssdc.asi.it/SED/} and are shown in gray. The {\it Swift UVOT/XRT} data from different states of the source are shown together with the {\it NuSTAR} spectrum. In the average state (pink), the 0.3-10 keV X-ray photon index is $\geq2.0$ which softens to $2.56 \pm 0.028$ in the 3-50 keV band. In the bright state (cyan), the photon index is $2.07\pm0.03$ but it can be as hard as $1.60 \pm 0.05$ on MJD 58489.1 (dark blue circle). In the \gray band, in addition to the \gray spectra averaged over $\sim11.7$ years, the spectrum in the period of the hardest \gray emission ($1.39 \pm 0.16$) is shown with a bow-tie. Even though the \gray light curve reveals periods when the flux moderately increases, their duration ($\sim$ 20-30 days) and the low amplitude ($\sim1.5-2.0$) do not impact the averaged \gray spectra in Fig. \ref{sed}. To demonstrate this, the \gray data analysis was performed limiting the time up to MJD 58500 which resulted in $(1.86\pm0.11)\times10^{-8}{\rm photon\;cm^{-2}\;s^{-1}}$ and $1.71\pm0.02$, in agreement with the results obtained for $\sim11.7$ years. The VHE \gray data are from the {\it VERITAS} observations after EBL correction \citep{2009ApJ...695.1370A}.\\
The multiwavelength SED in Fig. \ref{sed} shows dramatic changes in the spectrum of \es, especially in the X-ray band. In the quiescent state, the {\it NuSTAR} spectrum is a continuation of that of {\it XRT} and they together constrain the HE tail of the synchrotron component. Moreover, in quiescent state, the synchrotron and IC peaks have similar luminosity but unlike the HE peak which is relatively stable, the X-ray flux increases substantially (cyan data) and the low energy peak luminosity clearly dominates over that at HEs. In the active state, apart from the flux increase, the X-ray spectrum also hardened, shifting towards HEs (light blue data). Such an X-ray spectrum cannot be associated with the synchrotron component dominating at lower energies, being most likely due to another component. For example, this hard X-ray emission can be due to Comptonization of disc photons in the jet \citep{2007MNRAS.375..417C}. However, the origin of this X-ray emission cannot be investigated because of the absence of contemporaneous multiwavelength data. This change of X-ray spectrum will be further discussed in Section \ref{sec5}.
The HE \gray data clearly demonstrate that the peak of the second component is above $10^{11}-10^{12}$ eV which cannot be constrained even when VHE \gray data are considered because of the large uncertainty in the measurement of the VHE \gray photon index ($1.86 \pm 0.37$). However, independently of the VHE \gray data, the constraint on the synchrotron spectrum and hence on the distribution of emitting electrons will allow to shape the second component.\\
The SED of \es in Fig. \ref{sed} is modeled in order to gain further insight of the physical processes at work in its jet. The broadband spectrum of \es in the quiescent state is modeled within a simple one zone leptonic scenario as a large amount of data is available. In this model, the low energy data are interpreted by synchrotron emission of relativistic electrons while the HE component - as SSC radiation from a compact emitting region \citep{1996ApJ...461..657B, 1985A&A...146..204G, 1985ApJ...298..114M, 1999MNRAS.306..551C}. It is assumed that this emission region is a spherical blob moving relativistically along the axis of the jet with a Lorentz factor of $\Gamma$ and because of this, the emission will be strongly Doppler boosted in the observer frame by a factor of $\delta=(\Gamma (1-\beta cos \theta))^{-1}$ where $\theta$ is the angle between the direction of observation and the axis of the jet. For a small viewing angle $\Gamma\simeq\delta$. It is assumed that the blob is filled with an electron (or electron/positron) population in an isotropic magnetic field. For the electron energy distribution we consider a broken power-law (BPL) function in the form of
\begin{equation}
      N_e(\gamma)=
      \begin{cases}
	N_0 \left(\gamma\right)^{-\alpha} & \gamma_\mathrm{min} \leqslant \gamma \leqslant \gamma_\mathrm{br} \\
	N_0 \left(\gamma_{br}\right)^{\alpha_1-\alpha} \left(\gamma\right)^{-\alpha_1} & \gamma_\mathrm{br} \leqslant \gamma
      \end{cases}
      \label{BPL}
      \end{equation}
where $N_{0}$ defines the total electron energy $U_{\rm e}=\int_{\rm \gamma_{min}}^{\rm \gamma_{\rm max}}\gamma\: N_e(\gamma)d\gamma$, $\alpha$ and $\alpha_{1}$ are the low and high indexes of electrons, correspondingly below and above the break energy $\gamma_{\rm br}$. $\gamma_{\rm min}$ is the minimum energy of electrons in the jet frame. The electron distribution given by Eq. \ref{BPL} is a result of injection and cooling of particles \citep{1962SvA.....6..317K}. As an alternative, a power-law with an exponential cut-off (PLEC) distribution of particles is assumed:
\begin{equation}
      N_e(\gamma)= N_0 \gamma^{-\alpha}\: exp(-\gamma/\gamma_{\rm cut}) 
      \label{PLC}
      \end{equation}
where $\gamma_{\rm cut}$ is the highest energy cut-off in the electron spectrum. This electron distribution is naturally formed when the acceleration is limited by cooling or dynamical time scales (e.g., \citet{2013ApJ...765..122Y, 2018PASP..130h3001Z, 2020A&A...635A..25S, 2017MNRAS.464.4875B}). In the next section, time dependent formation of these spectra is discussed in the context of particle acceleration and cooling.\\
The emitting region is characterized by the electron energy distribution ($\alpha$, $\alpha_1$, $\gamma_{\rm br/cut}$), magnetic field, Doppler boosting and its size. The upper limit on the size of the emission region can be derived from the relation $R\leq\delta\:c\:t_{\rm var}$, where the variability time $t_{\rm var}$ can be inferred from the \gray light curve when the flare rise or decay time can be estimated. In the HE \gray band, \es did not show prominent flares, while in the VHE \gray band, the flux doubling time was observed to be $t_{\rm var}\leq 1$ day \citep{2010ApJ...709L.163A} which limits the emission region size by $R\leq2.19\times10^{15}\times\delta\: {\rm cm}$. Assuming a typical value for $\delta=25$, the emission region size would be $R\simeq5.5\times10^{16}\: {\rm cm}$. Also, following \citet{2010MNRAS.401..973R}, for the same  $R\simeq5.5\times10^{16}\: {\rm cm}$ the data is modeled considering $\delta=80$.\\
In the fit, the model free parameters and their uncertainties are estimated using the Markov chain Monte Carlo (MCMC) method in two approaches. Initially, the spectral model parameters have been derived through MCMC sampling of their likelihood distributions using a modified version of {\it naima} package \citep[e.g.,][]{2017MNRAS.470.2861S, 2018ApJ...863..114G}. Then, the fitting is done with the open source package {\it JetSet} \citep{2006A&A...448..861M, 2011ApJ...739...66T, 2009A&A...501..879T}, initially optimizing the parameter space with {\it Minuit} minimizer, then applying MCMC sampling centered on the best fit values. Both methods produce similar results.  
\subsection{Broadband SED fitting}
\begin{figure}
   \centering
   \includegraphics[width=0.48 \textwidth]{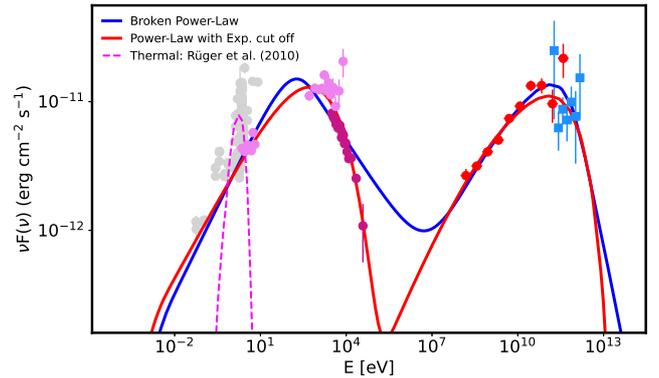}
      \caption{Broadband SED of \es modeled using a one-zone model. The red curve shows the emission assuming a power-law with an exponential cutoff distribution for the emitting electrons while that using a broken power-law is in blue.}
         \label{sed_model}
   \end{figure}
The SED modeling results are shown in Fig. \ref{sed_model} with the corresponding parameters listed in Table \ref{params}. In addition to the synchrotron/SSC component, a thermal component (a blackbody with a temperature of $T = 4500$ K) from \citet{2010MNRAS.401..973R} is shown in magenta. This corresponds to the thermal emission of the host galaxy, showing that the {\it UVOT} data in the average state are likely dominated by the nonthermal emission from the jet rather than by thermal emission from the host galaxy. It can be seen that the data up to the X-ray band (including {\it NuSTAR}) are explained by synchrotron radiation of electrons. The SSC component is dominant above $\sim1$ MeV and it describes the data up to the VHE \gray band. When the BPL distribution of electrons is considered (blue line in Fig. \ref{sed_model}), the data can be modeled when $\alpha=2.09\pm0.06$ changes to $3.67\pm0.10$ at the break energy of $(1.72\pm0.31)\times10^{5}$. So, the index change is significantly different from that expected when the electrons undergo radiative losses effectively ($\Delta \alpha=1$). The archival data allows to put a limit on the $\gamma_{\rm min}$ to be $(5.69\pm0.05)\times10^{2}$: for larger $\gamma_{\rm min}$ the synchrotron component will decline in the low energy band which is in disagreement with the observed data. The magnetic field is estimated to be $(1.58\pm0.21)\times10^{-2}$ G and the emission region is particle-dominated with an equipartition ratio between the particle energy density $U_{\rm e}$ and magnetic field energy density $U_{\rm B}$ of $U_{\rm e}/U_{\rm B}\simeq2.1\times10^2$.\\
The modeling with a PLEC distribution of electrons better explains the SED (red line in Fig. \ref{sed_model}); the goodness of fit (reduced $\chi$ square) is $\chi^2=0.90$. The power-law index is $2.19\pm0.04$ not substantially different from that estimated in the previous case. The synchrotron peak and {\it NuSTAR} spectrum allowed to measure the cut-off energy with a high accuracy, $\gamma_{\rm cut}=(4.73\pm0.34)\times10^{5}$, which in its turn constrains the HE component which decreases above $\sim10^{12}$ eV.  A relatively high value of the minimal energy of the radiating electrons, $\gamma_{\rm min}=(4.55\pm0.04)\times10^{2}$ is obtained which is not exceptional for blazar modeling, and high $\gamma_{\rm min}$ is often used to describe the emission from HSPs \citep[e.g.,][]{2011ApJ...736..131A,2018ApJ...862...41A, 2020arXiv200200129M}. The magnetic field is mostly constrained by fitting the low energy component; the synchrotron component depends on the product of $B$ and $N_{e}$, so $B=(1.55\pm0.09)\times10^{-2}$ G is the same as in the previous case. Again, the electron energy density is higher than that of the magnetic field, $U_{\rm e}/U_{\rm B}\simeq2.9\times10^2$.\\
The modeling parameters when $\delta=80$ are given in Table \ref{sed_model} in the brackets. As compared with the modeling when $\delta=25$, a noticeable difference in this case is that the electron distribution is with a softer power-law index ($\alpha=2.31$ and $2.43$) and the magnetic field is lower ($\simeq(1.6-1.7)\times10^{-3}$ G). As the peak of the low energy component is well defined by the data, when increasing the Doppler factor (and hence the luminosity), a lower magnetic field would be required to explain the same data. This, in its turn increases the particle dominance and now $U_{\rm e}/U_{\rm B}>10^4$.\\
The jet power in the form of magnetic field and particles is given in Table \ref{params}. The luminosities have been computed for a pure electron/positron jet, since the proton content is not well known, and these can be considered as the lower limit. The absolute jet power ($L_{\rm jet}\simeq8\times10^{43}{\rm erg\:s^{-1}}$) is significantly below the Eddington luminosity for a $5.6\times10^8\:M_\odot$ black hole mass ($L_{\rm Edd}=7.3\times10^{46}\:{\rm erg\:s^{-1}}$) estimated from the properties of the host galaxy in the optical band \citep{2010MNRAS.401..973R}.
\begin{table}
\caption{\label{params} Parameters of the models in Fig. \ref{sed_model}.}
\centering
\begin{tabular}{lcc}
\hline\hline
&PLEC $(\delta=80)$ &BPL $(\delta=80)$\\
\hline
$\alpha$           & $2.19\pm0.04$ $(2.31\pm0.03)$ &$2.09\pm0.06$ $(2.43\pm0.02)$\\
$\alpha_1$           & -- &$3.67\pm0.10$ $(4.37\pm0.15)$\\
$\gamma_{\rm min}\times10^2$           & $4.55\pm0.04$ $(5.07\pm0.10)$ &$5.69\pm0.05$ $(1.67\pm0.03)$ \\
$\gamma_{\rm cut/break}\times10^{5}$           & $4.73\pm0.34$ $(9.57\pm0.82)$ &$1.72\pm0.31$ $(7.47\pm0.79)$\\
$B[{\rm G}]\times10^{-2}$           & $1.53\pm0.09$ $(0.16\pm0.07)$ &$1.58\pm0.21$ $(0.17\pm0.01)$\\
$U_{\rm e}[{\rm erg\:cm^{-3}}]$           & $2.68\times10^{-3}$ $(2.24\times10^{-3})$  &$2.15\times10^{-3}$ $(3.77\times10^{-3})$ \\
$U_{\rm B}[{\rm erg\:cm^{-3}}]$           & $9.31\times10^{-6}$ $(9.92\times10^{-8})$  &$9.96\times10^{-6}$ $(1.13\times10^{-7})$ \\
$L_{\rm e} [{\rm erg\:s^{-1}}] $           & $7.64\times10^{43}$ $(6.39\times10^{43})$ &$6.11\times10^{43}$ $(1.07\times10^{44})$ \\
$L_{\rm B} [{\rm erg\:s^{-1}}] $           & $2.65\times10^{41}$ $(2.83\times10^{39})$   &$2.84\times10^{41}$ $(3.23\times10^{39})$\\
\hline
\end{tabular}
\end{table}   
\section{Time dependent formation of electron spectrum: Electron cooling}\label{sec4}
\begin{figure}
   \centering
   \includegraphics[width=0.48 \textwidth]{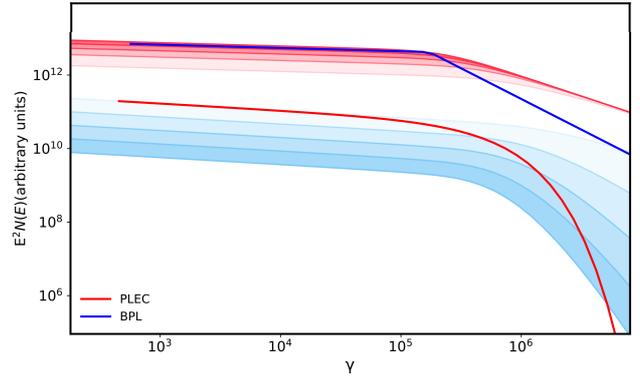}
      \caption{Time evolution of electron spectra considering different initial injection rates. The BPL and PLEC spectra in arbitrary units are shown in blue and red, respectively.}
         \label{electron}
   \end{figure}

The multiwavelength modeling presented in the previous section allowed to put a constraint on the parameters of emitting electrons. These parameters contain valuable information on the processes taking place in the jet of \es. For example, the power-law index of the emitting electrons mostly defined by the acceleration mechanisms could be used to test the process by which the particles gain energy. On the other hand, the break or cutoff energy allows to evaluate the electron radiation cooling or dynamical timescales, which helps to understand the particle interaction processes. Thus, the parameters reported in Table \ref{params} can be used for further exploring the physics of \es jet.\\
The spectra given in Eqs. \ref{BPL} and \ref{PLC} are ad-hoc assumption of emitting particles used for modeling the SED. However, the formation of the particle spectrum is governed by the injection and cooling of electrons. To calculate the temporal evolution of the electron distribution [$N_{\rm e}(\gamma,t)$], it is necessary to solve integro-differential equations, describing the losses and injection of relativistic electrons in the emitting region \citep{1962SvA.....6..317K}. In this case the kinetic equation has the following form
\begin{equation}
\frac{\partial N_{e}(\gamma,t)}{\partial t}=\frac{\partial}{\partial\gamma}\left(\dot{\gamma} N_{e}(\gamma,t)\right)-\frac{N_{e}(\gamma, t)}{t_{\rm esc}}
                          +Q(\gamma,t)\,,
\label{kinetic}
\end{equation}
where $\dot{\gamma}=d \gamma/dt$ is the radiation loss term, $t_{\rm esc}$ is the characteristic escape time and $Q(\gamma,t)$ is the rate of electron injection. The emitting region electrons can loose energy by synchrotron and SSC processes, so $\dot{\gamma}=-4/3\:\sigma_{\rm T}\:c U^\prime\:\gamma^2$ where $U^\prime$ is either the energy density of the magnetic field ($U_{B}^\prime=B^2/8\pi$) or the density of synchrotron radiation ($U^\prime_{\rm s}$). The latter can be approximated by $U^\prime_{\rm s}\simeq L_{\rm syn}^{\prime}/4 \pi\:R^2\:c$ where $L_{\rm syn}^{\prime}=\int \partial L_{\rm \nu, syn}/\partial \nu\:d\nu$ is the total synchrotron luminosity in the jet frame. The modeling shows that $L_{\rm syn}=5.74\times10^{41}\:{\rm erg\:s^{-1}}$ so $U^\prime_{\rm s}\simeq L_{\rm syn}/4 \pi\:R^2\:c\:\delta^3\simeq3.2\times10^{-8}\:{\rm erg\:cm^{-3}}$ which is significantly less than $U_{B}^\prime=9.31\times10^{-6}\:{\rm erg\:cm^{-3}}$. This implies that the electrons are predominantly cooled through interaction with the magnetic field. However, usually $U^\prime_{\rm s}$ should be taken into account, considering the non-linear effects in the particle cooling which is particularly strong when the emission is produced in a very compact region $\sim10^{15}\:{\rm cm}$. In this case, also due to the narrow distribution of synchtron photons (low energy component in Fig. \ref{sed_model}), its density is lower than that of the magnetic field. Accordingly, the radiation loss term in Eq. \ref{kinetic} is replaced by pure synchrotron cooling.\\
In the case of no escape ($t_{\rm esc}\rightarrow\infty$), that is all the particles cool inside the emitting region, a BPL spectrum of the electrons will be formed when the power-law index changes as $\alpha_1-\alpha=1$. The break energy will be defined by equating the cooling time with the evolution time of the system. In Fig. \ref{electron} the evolution of the spectrum when the particles are constantly injected ($t_{\rm inj}>>t_{\rm cool}$) into the emitting region with $Q(\gamma)\sim \gamma^{-2.09}$ is shown for different dynamical time scales; the red gradient shows the spectrum with increasing time. After the system evolves up to $\sim1.80\times10^7$ sec, a break at $1.72\times10^5$ will be formed in the spectrum; for shorter times the break is at higher energies. However, as expected, the transition at the break energy is smooth ($2.09\rightarrow 3.09$) which cannot explain the estimated electron spectrum obtained from the data modeling (blue spectrum in Fig. \ref{electron}). A steep electron spectrum is required after the break to explain the X-ray data; the {\it NuSTAR} spectrum completely defines the power-law index of the electrons after the break to be $2\times\Gamma_{\rm X-ray}-1=4.12$. It means that the power law index of electrons before the break should be $3.12$ but in that case their \gray emission will be nearly flat $\sim2.06$ in disagreement with the data ($\sim E_{\gamma}^{-1.76}$). The electron power-law indexes before and after the break are rather well defined by the X-ray and \gray data, so when changing the values of $\alpha$ and $\alpha_1$ reported in Table \ref{params} to get a cooling break, the data will not be satisfactorily reproduced. When assuming a constant injection of particles with their escape timescale depending on the energy ($\sim \gamma^\epsilon$), more gradual transition will be achieved at the cooling break when $\epsilon\neq0$, but again a component with a softer spectrum cannot be formed. Most likely, this break is due to the characteristics of the acceleration processes and for an unknown reason the change in the electrons spectrum is $\Delta \alpha>1$. Or alternatively, the inhomogeneities in the emitting region could also cause a stronger change in the emitting electron spectrum, which might produce BPL spectrum of electrons with $\Delta \alpha>1$ \citep{2009ApJ...703..662R}.\\
In the case the electrons can escape from the emitting region, a standard BPL spectrum will be formed again, only the break will correspond to the electron energy at which the escape and cooling timescales are equal ($t_{\rm esc}=3 m_e\:c/4 \sigma_{\rm T} U_{\rm B}\:\gamma_{\rm br}$). For example, a BPL spectrum at $\gamma_{\rm br}=1.72\times10^5$ can be formed when $t_{\rm esc}=9.83R/c$. However, the transition at the break is again not sharp enough to explain the observed data. \\
Alternative to BPL, a PLEC spectrum can be formed as a result of time averaging of the injected particle spectrum, i.e., after the abrupt power-law injection of the particles ($t_{\rm inj}<t_{\rm cool}$) they start to cool in the emitting region. In time, the HE tail of the particle distribution steepens and a cut-off will be formed. In order to demonstrate this, it is assumed that the $Q(\gamma)\sim \gamma^{-2.19}$ injection of the particles stops at $R/10\:c$ and then the electron distribution evolution up to $10\:R/c$ is followed by setting $t_{\rm esc}=1.5 R/c$ and $B=1.53\times10^{-2}$ G. The blue gradient in Fig. \ref{electron} corresponds to the electron spectrum calculated for different time intervals. Initially, only the HE electrons ($\gamma>10^6$) cool or escape the region, declining the injected electron spectrum only at higher energies. Then, with the time the cut-off energy moves to lower energies and after $\sim3-4\:R/c$ the break is at the same level as that estimated from the data modeling ($1.58\times10^5$). In principle, by changing the parameters (injection and escape times, etc.), it is possible to satisfactorily reproduce the PLEC spectrum of electrons with the parameters given in Table \ref{params}. It should be noted that such an exponential cut-off will be also formed in the case of an episodic injection with an energy dependent escape.
\section{Discussions and Conclusions}\label{sec5}
\begin{figure}
   \centering
   \includegraphics[width=0.48 \textwidth]{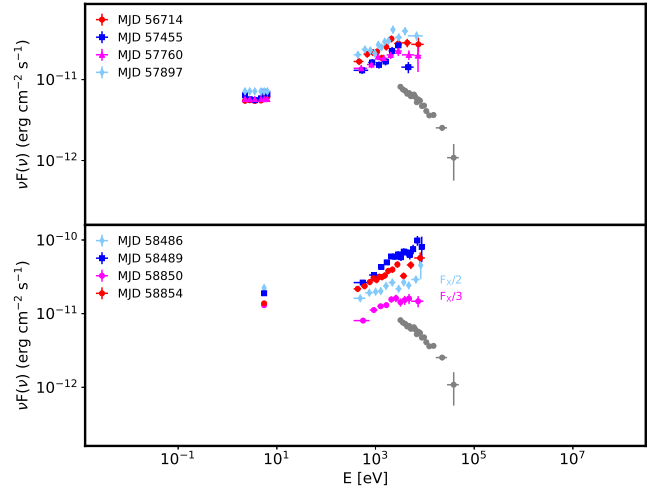}
      \caption{{\it Swift UVOT} and {\it XRT} spectra of \es during the extreme X-ray emission period as compared with that of {\it NuSTAR} (gray).}
         \label{hard}
   \end{figure}

The HSP blazars having the second peak in the SED toward the higher energies are extremely interesting sources for HE and VHE \gray observations. For HSPs, the X-ray spectrum limits the synchrotron component below keV/MeV band and the \gray emission is due to IC scattering. In the \gray band, HSPs appear with a moderately hard photon index ($<2.0$) with a mean of $1.81\pm0.08$ \citep{2020ApJ...892..105A}, so the emitting particles are accelerated to much higher energies as compared with LSPs or ISPs. This implies the multiwavelength observations have potential not only for investigation of the emission features in different bands but also for testing various acceleration and emission scenarios in the blazar jets.\\ 
The \gray emission from \es is investigated during more than eleven years, from August 2008 to April 2020. Its long time-averaged \gray spectrum is hard with a photon index of $1.71\pm0.02$ and with a flux of $(1.89\pm0.09)\times10^{-8}\:{\rm photon\:cm^{-2}\:s^{-1}}$. The emission extends up to $\sim600$ GeV with a detection significance of $77.2\sigma$. The sub-GeV source photons are relatively less, which prevents detailed variability studies in short time scales. Although the adaptively binned light curve shows several periods (e.g., on MJD $58869.84\pm13.82$, $58594.42\pm12.95$ and $56080.57\pm23.54$) when the flux moderately increased as compared to its average level. During these periods the data accumulation of 20-30 days is enough to reach the required uncertainty of 20 \% (usually $>(100-150)$ days are required). The photon index of the source is relatively constant with the hardest and softest values being $1.44\pm0.11$ and $2.01\pm0.17$, respectively. The hard emission observed in MJD $58272\pm10$ extends up to $\sim200$ GeV with a flux of $(5.09\pm1.47)\times10^{-7}{\rm photon\;cm^{-2}\;s^{-1}}$ and photon index of $1.39\pm0.16$ above $1$ GeV. In general, the spectrum measured by \fermi agrees well with that non-simultaneously measured by {\it VERITAS} at VHE \grays (after EBL correction).\\
As a bright source in the X-ray band, \es shows interesting features in the 0.3-10 keV band. First of all, an X-ray flux amplification in different observations is found with a highest X-ray flux of $(1.13\pm0.02)\times10^{-10}\:{\rm erg\:cm^{-2}\:s^{-1}}$. In the hard X-ray band, as observed by {\it NuSTAR}, the spectrum is soft with $2.56\pm0.028$ photon index. In the quiescent state, the {\it Swift XRT} and {\it NuSTAR} measured spectra are in a good agreement, allowing to measure the \es spectrum in the broad band of 0.3-50 keV. Yet, the {\it Swift XRT} observations reveal interesting modification of the X-ray photon index in some observations. It is mostly above $2.0$, as expected for HPS, but there are periods when the photon index hardens to $\leq1.80$. In Fig. \ref{hard} eight periods when such hard photon index is observed are presented together with {\it Swift UVOT} and {\it NuSTAR} data where the modification of X-ray spectra is evident. In the quiescent state, the nearly flat spectrum measured by {\it Swift XRT} and the soft X-ray photon index obtain by {\it NuSTAR} clearly imply that the synchrotron peak is $<10^{17}$ Hz. However, in the specific periods shown in Fig. \ref{hard}, the position of the synchrotron peak moves above $10^{17}$ Hz and \es shows characteristics similar to EHSPs. This peak frequency shift is more evident and drastic for the periods highlighted in the lower panel of Fig. \ref{hard}. It should be noted that the optical/UV flux did not change substantially, but not always all the filters are available for detailed investigation. During the observations on MJD 57760 and 57897 the {\it UVOT} data could either correspond to the peak of the host galaxy emission (thus appear with a nearly flat spectrum) or be due to the synchrotron emission from the jet electrons but from a different component that produces hard X-ray emission. However, due to the lack of data no definite conclusion can be drawn. Such temporary extreme behavior of HPSs has already been observed for Mrk 501 \citep{1998ApJ...492L..17P} and 1ES 2344+514 \citep{2000MNRAS.317..743G}. In these extreme periods, the high counts ($\geq 280$) allowed precise estimation of the photon index which was $1.60\pm0.05$ and $1.70\pm0.07$ on MJD 58489 and 58854, respectively, and being within $1.72-1.84$ for the other periods. Also, the {\it MAGIC} and {\it VERITAS} observations of \es reveal an exceptional hard photon index in the TeV band ($<2.0$), though not simultaneous with the X-ray observations, which shows \es might have features similar to those of BL Lacs extreme in the \gray band. Along with these features also in the \gray band \es does not show a short time scale variability compatible with the behaviour of EHSPs. Yet, in the \gray band some of VHE photons from the direction of \es were observed around those extreme X-ray periods. For example, VHE events with $E_{\gamma}=292.0$, $278.1$ and $150.3$ GeV were observed on MJD 57453.7, 58498.3 and 57429.6, respectively, within a circle of $<0.1^\circ$ around \es and with a $>3.0\sigma$ probability to be associated with it. These periods overlap with some highlighted in Fig. \ref{hard}. The are also $>100$ GeV events emitted close to the periods in Fig. \ref{hard} which come from a bit larger circular region or with a smaller probability of association with \es.  However, considering \es is the only source in the ROI with emission extending above tens of GeV, these photons are most likely  also associated with it. This shows that during the extreme X-ray emission periods of \es, also GeV/TeV photons were efficiently produced which hints at possible transition of \es to an extreme BL Lac from the viewpoint of both synchrotron peak and VHE \gray photon index. It is expected that such extreme periods are hidden in the spectrum of HSP blazars and sometimes the transition of the synchrotron peak to higher frequencies is possible.\\
The quiescent state SED is modeled within a one-zone leptonic scenario. The synchrotron/SSC model well explains the observed data and can reproduce both low and HE peaks. The low energy photons with a peak at $\nu_{\rm sync}\simeq7\times10^{16}$ Hz, well constrained by {\it XRT} and {\it NuSTAR} data, are IC up scattered to higher energies $4/3\:\gamma_{\rm cut/br}^2\:\nu_{\rm sync}\simeq10^{27}$ Hz, explaining the second peak. The derived magnetic field in the jet emitting region is $B=1.5\times10^{-2}$ G for $R\simeq5.5\times10^{16}\: {\rm cm}$, the system being slightly particle dominated $U_{\rm e}/U_{\rm B}\simeq290$ which is in agreement with the SED modeling of other HSPs; usually within the leptonic scenario the HSP SEDs can be modeled when the emitting region is by far out of equipartition.
In this case the equipartition is between the magnetic field and nonthermal electron energy density, and it would break when considering the jet protons, the content of which is unknown. The energy density of electrons strongly depends on $\gamma_{\rm  min}\sim500$ which in this case should be considered as an upper limit; in the case when $\leq\gamma_{\rm  min}$ the SED can be still described well but when $>\gamma_{\rm  min}$, the model starts to drop in disagreement with the observed data. When the modeling infers an extremely out-of-equipartition condition in the jet, additional jet power is required which is however limited by the Eddington accretion rate \citep[e.g.,][]{2015ApJ...809..174D}. As an alternative, in highly magnetized environments the combined lepto-hadronic modeling would allow to choose parameters and explain the data when the system is close to the equipartition condition \citep[e.g., modeling of 3C 279][]{2016ApJ...832...17B}.\\
The one-zone SSC model was also used to model the \es SED in the previous studies \citep[e.g.,][]{2010MNRAS.401..973R, 2010A&A...515A..18W, 2018MNRAS.477.4257C, 2019MNRAS.489.5076S}. For example, in \citet{2010MNRAS.401..973R}, using a time dependent code taking into account cooling of electrons and time evolution of photons the SED of \es was modeled for the electron distribution with $\alpha=2.1$  and $\gamma_{\rm cut}=5\times10^5$. Or in \citet{2019MNRAS.489.5076S}, a log-parabolic distribution function for the electrons can reproduce the observed data for index of $1.8$ and curvature parameter of $0.5$. Thus, in the previous modelings even if different assumptions were made for the emitting region size and Doppler boosting, the parameters obtained for the emitting electrons do not differ significantly from those obtained here. The energetics (e.g., luminosity or energy density) is different in all modelings as it depends on $R$ and $\delta$ the values of which were different. However, like in all the models, here too $U_{\rm e}/U_{\rm B}>10$. As an alternative to leptonic models, the HE bump of \es can be modeled by the proton synchrotron component when the magnetic field in the region is $B=(3.4-454)$ G and the protons are accelerated up to ultra high energies $2.4\times10^{19}$ eV and $(U_{\rm e}+U_{\rm p})/U_{\rm B}=2.2\times10^{-2}$ \citep{2015MNRAS.448..910C}.\\
The BPL and PLEC electron spectra were used to model the SED. In general, the spectra of electrons is controlled by several timescales, namely, the cooling time, the injection duration and the escape time. In the continuous injection of particles and in the absence of escape, the traditional cooling break formed in the electron spectrum can not explain the tail of the high energy component which requires much stepper decrease: the cooled electron spectrum above $1.72\times10^5$ will exceed that obtained from the modeling. The change in the electron spectrum is most likely caused by the nature of the acceleration process. Alternatively, \citet{2009ApJ...703..662R} showed that when synchrotron losses are dominating, spectral breaks in the electrons spectrum differing from $0.5$ can be naturally formed in inhomogeneous sources. These can be straightforwardly applied to pulsar wind nebulae or knots in large-scale jets, but may be applied also wherever bulk flows of relativistic material are involved, as in the case of relativistic jets. On the other hand, the time limited power-law injection of electrons which cool in the emitting region will stabilize and form a cutoff in the electron distribution in time. The value of cut-off energy depends on the time for which the system evolves and $\gamma_{\rm cut}=(4.73 \pm 0.34)\times10^5$ obtained from the SED modeling can be naturally obtained. The required time for dynamical evolution of the system is 3-4 $R/c$ which is in agreement with the absence of flaring activities in \gray band in short time scales. In principle, more complicated scenarios for the formation of a curved spectrum are possible, but here a simplified scenario when the curvature is caused by the injection/cooling or energy-independent escape from the emitting region, is discussed.\\
As a powerful HSP, \es has always been monitored in various bands; it is still debatable whether \es is a normal HSP or an extreme blazar. Some of the {\it Swift XRT} observations analyzed here appeared with an extremely hard photon index of $\leq1.8$ shifting the X-ray spectrum toward higher frequencies, making \es an episodic extreme synchrotron blazar. It should be noted that a smooth transition within the blazar classification is emerging in some observations, e.g., a classical FSRQ shows a BL Lac features during the flares \citep[e.g.,][]{2014MNRAS.445.4316C, 2012MNRAS.421.1764S, 2002ApJ...571..226C, 2012MNRAS.420.2899G} or HSPs appear as extreme blazars \citep[e.g.,][]{1998ApJ...492L..17P, 2000MNRAS.317..743G, 1999APh....11...11G}. Identification of such hidden periods when HSPs are in an extreme emission state with a large multifrequency coverage can be crucial for understanding the physics of extreme blazars and investigation of the changes in the jet parameters causing their extreme behaviour.

\section*{Acknowledgements}
I thank the anonymous referee for constructive comments. This work was supported by the RA MoESCS Committee of Science, in the frames of the research project No 18T-1C335. This work used resources from the ASNET cloud and the EGI infrastructure with the dedicated support of CESGA (Spain).
\section*{Data availability}
The data underlying this article will be shared on reasonable request to the corresponding author.



\bibliographystyle{mnras}
\bibliography{biblio} 

\end{document}